\documentclass{elsart}

\usepackage{cite}

\usepackage{graphicx}

\begin{document}

\begin{frontmatter}

\title{On some approximate methods for nonlinear models}

\author{Francisco M. Fern\'{a}ndez \thanksref{FMF}}

\address{INIFTA (UNLP,CCT La Plata-CONICET), Divisi\'{o}n Qu\'{i}mica Te\'{o}rica,
Diag. 113 y 64 (S/N), Sucursal 4, Casilla de Correo 16,
1900 La Plata, Argentina}

\thanks[FMF]{e--mail: fernande@quimica.unlp.edu.ar}

\begin{abstract}
We show that recent applications of the homotopy perturbation method
the Adomian decomposition method and the variational iteration method
are completely useless for the treatment of nonlinear problems.

\end{abstract}

\end{frontmatter}

\section{Introduction}
In a series of papers we have shown that some particular applications of
some variational and perturbation approaches (VAPA) like, for example,
the homotopy perturbation method (HPM), homotopy
analysis method (HAM), Adomian decomposition method (ADM) and variational
iteration method (VIM) are utterly
useless for the study of nonlinear systems, even for the simplest
models\cite{F07,F08b,F08,F08c,F08d,F08e,F08f}. Unfortunately some journals
do not accept criticisms of the papers they publish.

The purpose of this paper is the analysis of some recent applications of VAPA to
simple models of nonlinear phenomena published in this journal. In
Sec.~\ref{sec:HomoRicc} we discuss the application of HPM to an exactly solvable
Riccati equation. In Sec.~\ref{sec:Prey-predator} we analyse the application of
the same approach to an exactly--solvable model for the interaction between a
prey and a predator. In Sec.~\ref{sec:SIR} we discuss the application of other
VAPA to the simplest epidemic model. Finally, in Sec.~\ref{sec:conclusions} we
draw some conclusions
on the achievements and future of those VAPA.

\section{Homotopy perturbation method for a quadratic differential
equation} \label{sec:HomoRicc}

Lately, there has been great interest in the application of analytical
approximate methods to the solution of several problems of nonlinear dynamics%
\cite{BM05,BIK05,A06,CHA07,RDGP07,YE08}, some of which lead to time--power
series\cite{BM05,BIK05,A06,CHA07}. Some of the models considered by those
authors are suitable for the study of prey--predator interactions\cite
{BM05,BIK05,CHA07,RDGP07,YE08}.

It is curious that most of those authors resort to a power series
description of the dynamics of the system because it is known that such
approach is unable to provide a reasonable overall picture of the evolution
which is what really matters in the case of, for example, prey--predator
models. A time--power series is limited to a neighbourhood of the origin
(for example, initial species population) by its convergence radius that is
determined by the singularity closest to that point. Besides, most equations
for nonlinear dynamics generate singularities spontaneously that move when
the initial conditions change. We have already pointed out the limitations
of several of those approaches in earlier
communications\cite{F07,F08b,F08,F08c,F08d,F08e,F08f}.

The purpose of this section is to discuss some of those features of the
nonlinear dynamics by means of the simple Riccati equation solved by
Abbasbandy\cite{A06} by means of the HPM.

The model chosen by Abbasbandy\cite{A06}
\begin{equation}
\frac{dY(t)}{dt}=2Y(t)-Y(t)^{2}+1,\;Y(0)=0  \label{eq:Riccati}
\end{equation}
for the application of the HPM is useful for present discussion because the
Riccati equation can be solved exactly:
\begin{equation}
Y(t)=\frac{e^{2\sqrt{2}\,t}-1}{\left( \sqrt{2}-1\right) e^{2\sqrt{2}\,t}+%
\sqrt{2}+1}  \label{eq:solution_0}
\end{equation}
One clearly appreciates that there is a pole in the complex $t$--plane that
limits the convergence of the time--power series to $t<|t_{c}|$, where
\begin{equation}
|t_{c}|=\frac{\sqrt{2}}{4}\sqrt{4\left[ \ln \left( \sqrt{2}+1\right) \right]
^{2}+\pi ^{2}}\approx 1.274  \label{eq:tc_0}
\end{equation}
In other words, the HPM proposed by Abbasbandy\cite{A06} will be useless for
$t>|t_{c}|$ disregarding the order of the perturbation approach. In
particular, the resulting time--power series will not reveal the stationary
point $Y_{s}=\sqrt{2}+1$\cite{EBA04} to which the solution approaches as $%
t\rightarrow \infty $. Such stationary points are most important for any
physical description of a dynamical system\cite{BO78,S94}.

The solution to the Riccati equation (\ref{eq:Riccati}) with an arbitrary
initial condition $Y(0)=Y_{0}$ is
\begin{equation}
Y(t)=\frac{\left( \sqrt{2}+1\right) \left( Y_{0}+\sqrt{2}-1\right) e^{2\sqrt{%
2}\,t}+Y_{0}\left( \sqrt{2}-1\right) -1}{\left( Y_{0}+\sqrt{2}-1\right) e^{2%
\sqrt{2}\,t}-Y_{0}+\sqrt{2}+1}  \label{eq:solution_Y0}
\end{equation}
and the radius of convergence of the time--series is
\begin{equation}
|t_{c}|=\frac{\sqrt{2}}{4}\ln \left( \frac{Y_{0}-\sqrt{2}-1}{Y_{0}+\sqrt{2}-1%
}\right)  \label{eq:tc_Y0}
\end{equation}
Notice that the singularity closest to the origin of the complex $t$--plane
moves as the initial condition changes as mentioned above. The HPM performs
even more poorly if $Y_{0}>Y_{s}$; for example, if $Y_{0}=5$ then $%
|t_{c}|\approx 0.261$.

From the discussion above we clearly appreciate that the HPM proposed by
Abbasbandy\cite{A06} (and also the time--power series produced by any other
approach\cite{BM05,BIK05,CHA07}) is unable to reveal the main features of
nonlinear dynamical models.

Straightforward application of the ADM\cite
{BM05,EBA04} also produces a time--power series, however, El--Tawil et al%
\cite{EBA04} proposed a multistage ADM that
basically leads to the expansion of $Y(t+\Delta t)$ in $\Delta t$--power
series. The radius of convergence of this series is given by
\begin{eqnarray}
|\Delta t_{c}|(t)&=&\frac{\sqrt{2}}{4}\sqrt{4\ln \left( \sqrt{2}+1\right)
^{2}-8\sqrt{2}\ln \left( \sqrt{2}+1\right) t+8t^{2}+\pi ^{2}} \\ \nonumber
&>&\frac{\sqrt{2}\pi }{4}\approx 1.11  \label{eq:Dtc_0}
\end{eqnarray}
when $Y_{0}=0$. El--Tawil et al\cite{EBA04} chose $\Delta T\ll 1.11$ and
their multistage ADM yielded accurate results of
$Y(t)$ even for values of $t$ sufficiently large that $Y(t)\approx Y_{s}$.
However, this approach is more closely related to numerical integration
algorithms such as Runge--Kutta than to the analytical methods discussed
above.

Analytical approaches like the HPM and ADM that
lead to time--power series\cite{BM05,BIK05,A06,CHA07} are completely useless
for a serious study of nonlinear dynamics because they fail to provide the
overall features of the problem that any physical or ecological application
requires\cite{F07,F08b,F08,F08c,F08d,F08e,F08f}. Other approaches like the
VIM\cite{YE08} also yield expressions that are just valid in a
meaningless neighbourhood of origin\cite{F08}.

\section{Homotopy perturbation method for a prey--predator model}
\label{sec:Prey-predator}
In a recent paper Rafei et al\cite{RDGP07} proposed the application of the
HPM to the simplest model of prey--predator
interaction. After a rather tedious development of the main equations, their
particular implementation of the HPM leads to a series solution of the
population of the species. It is difficult to justify the application of the
HPM in this way because one obtains the same solution more easily and
straightforwardly by direct expansion of the nonlinear differential equation
in time--power series\cite{BIK05}. Besides, we have recently shown that the
time--power series developed by Chowdhury et al\cite{CHA07} by means of an
alternative implementation of the HPM completely fails to yield the main
features of the population dynamics\cite{F07}. The ADM also leads to a
time--series solutions although in this case they are
slightly different from the correct ones\cite{BM05}.

The purpose of this section is to analyze the results of Rafei et al\cite
{RDGP07} in terms of what one expects from an approach designed to solve
problems of population dynamics. The results of this analysis applies also
to the other methods already mentioned above\cite{BIK05,CHA07,BM05}.

Rafei et al\cite{RDGP07} chose the prey--predator system
\begin{eqnarray}
\frac{dx(t)}{dt} &=&x(t)[a-by(t)],\;a,b>0  \nonumber \\
\frac{dy(t)}{dt} &=&-y(t)[c-dx(t)],\;c,d>0  \label{eq:model}
\end{eqnarray}
where $x(t)$ and $y(t)$ are the populations of rabbits and foxes,
respectively. This nonlinear system (\ref{eq:model}) exhibits a saddle point
at $(x_{s},y_{s})=(0,0)$ and a center at $(x_{s},y_{s})=(c/d,a/b)$. Besides,
the populations obey the following curve in the $x-y$ plane:
\begin{equation}
\ln \left( x^{c}y^{a}\right) -dx-by=\ln \left( x_{0}^{c}y_{0}^{a}\right)
-dx_{0}-by_{0}  \label{eq:curve}
\end{equation}
where $x_{0}$ and $y_{0}$ are the initial populations at time $t=0$.

The HPM proposed by Rafei et al\cite{RDGP07} leads to time--series
expansions of the form
\begin{eqnarray}
x(t) &=&x_{0}+x_{0}(a-by_{0})t+\ldots   \nonumber \\
y(t) &=&y_{0}+y_{0}(dx_{0}-c)t+\ldots   \label{eq:t-series}
\end{eqnarray}
that are exactly the same as those obtained earlier by Biazar et al\cite
{BIK05}. Obviously, they are suitable about the point $(x_{0},y_{0})$ and
will not predict the main features of the population dynamics which is what
really matters in this field\cite{BO78}.

\begin{figure}[H]
\begin{center}
\includegraphics[width=9cm]{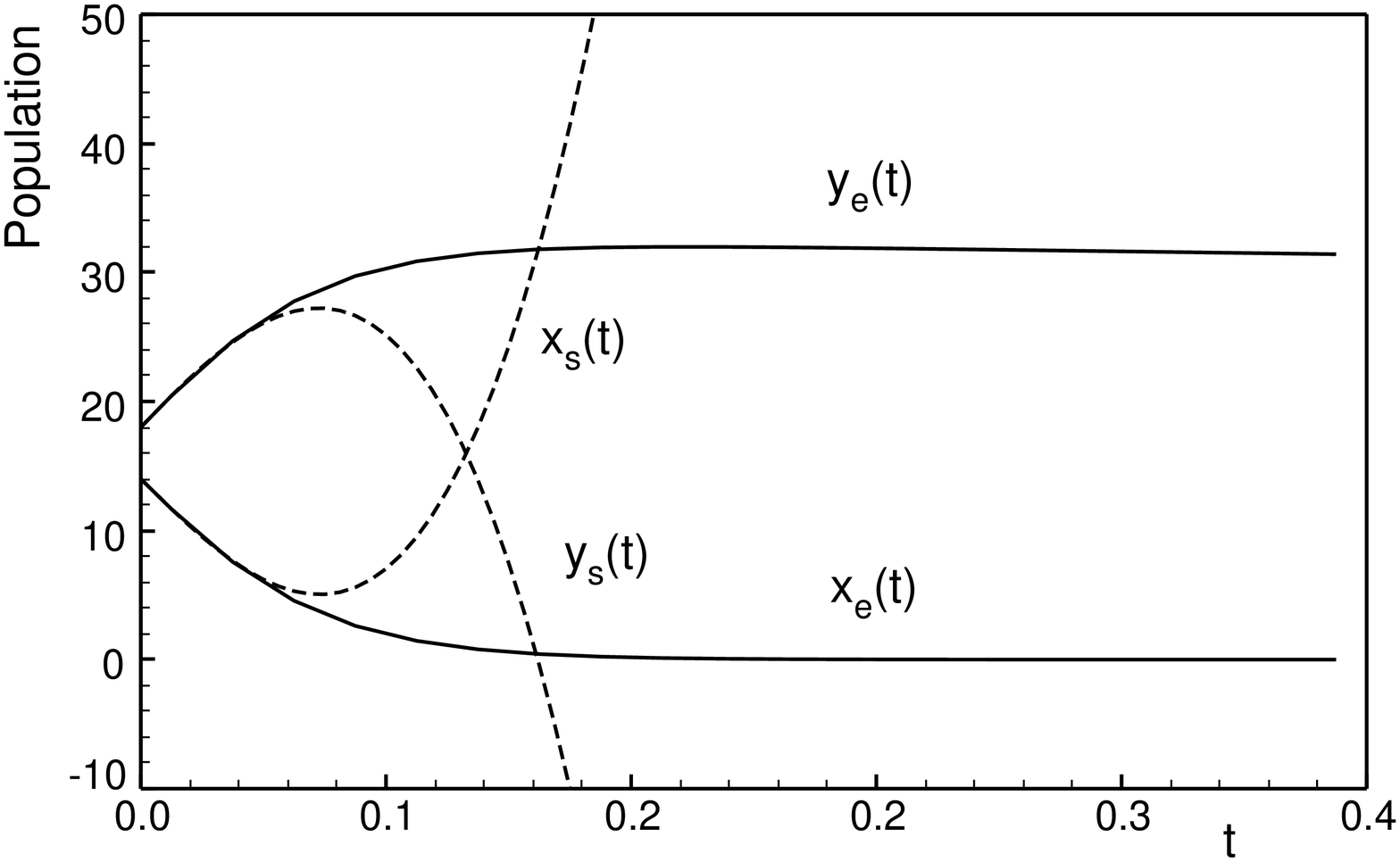}
\end{center}
\caption{Exact (solid, e) and series (dashed, s) populations for Case I}
\label{Fig:RDGP1}
\end{figure}

Fig.~\ref{Fig:RDGP1} shows the populations for Case I ($a=b=d=1$, $c=0.1$, $%
x_{0}=14$, $y_{0}=18$) in a time scale larger than that considered in the
earlier studies already mentioned above\cite{RDGP07,BIK05,BM05}. We clearly
appreciate that all those approaches predict a wrong behaviour of the
populations.

\begin{figure}[H]
\begin{center}
\includegraphics[width=9cm]{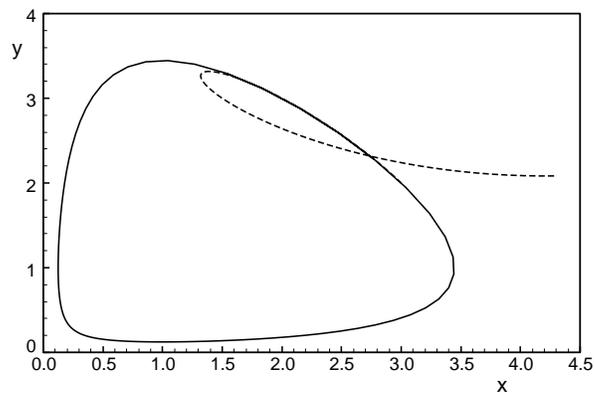}
\end{center}
\caption{Exact (solid) and approximate (dashed) populations for Case V
in the $x-y$ plane}
\label{Fig:RDGP2}
\end{figure}

Case V ($a=b=c=d=1$, $x_{0}=3$, $y_{0}=2$) is much more interesting. Fig.~%
\ref{Fig:RDGP2} shows the results in the plane $x-y$. The time series
predict a completely wrong behaviour and the resulting curve cross itself
which can never happen as everyone knows\cite{BO78}.

Present analysis clearly shows that the HPM\cite{RDGP07,CHA07},
the ADM\cite{BIK05} and the
straightforward time--series expansion\cite{BM05} are completely useless for
a reasonable prediction of the evolution of even the simplest prey--predator
systems.

\section{Simple epidemic model for non--fatal disease}\label{sec:SIR}

In a series of papers Biazar\cite{B06} and Rafei et al\cite{RDG07,RGD07}
discussed the application of VAPA to the simplest epidemic
model of a non--fatal disease, commonly called SIR and attributed to Kermack
and McKendrick\cite{KM27}. Those authors applied ADM\cite{B06},
VIM\cite{RGD07}, and HPM\cite{RGD07}, and obtained the Taylor
expansions of the number of susceptibles, infectives, and removed about the
initial time $t=0$. In those papers the authors verified that the three
methods yield exactly the same expansions, and, consequently, they showed
the same figures for the evolution of the epidemic.

However, Biazar\cite{B06} and Rafei et al\cite{RDG07,RGD07} did not try to
simulate any real epidemic situation and did not explain the reason for
choosing their particular values of the model parameters and initial
conditions.

The purpose of this section is to investigate to which extent the analytical
expressions derived by those authors are useful for a reasonable prediction
of the behaviour of an epidemic within the realm of the rather
oversimplified SIR model. We first answer some relevant
questions about the epidemic dynamics by means of the exact solution. Then
we choose particular model parameters and study if
the approximate analytical expressions proposed by Biazar\cite{B06} and
Rafei et al\cite{RDG07,RGD07} are suitable for answering those questions.
Finally, we give our opinion about the utility of those analytical expressions.

The nonlinear ordinary differential equations that predict the evolution of
the epidemic according to the SIR model are:
\begin{eqnarray}
\frac{dx}{dt} &=&-\beta xy  \nonumber \\
\frac{dy}{dt} &=&\beta xy-\gamma y  \nonumber \\
\frac{dz}{dt} &=&\gamma y  \label{eq:model_eqs}
\end{eqnarray}
where $x$ is the number of susceptibles, who do not have the disease but
could get it, $y$ is the number of infectives, who have the disease and can
transmit it to others, $z$ is the number of removed, who cannot get the
disease or transmit it, and $\beta ,\gamma >0$ are model parameters that
determine the epidemic evolution.

Biazar\cite{B06} proposed to solve the SIR equations (\ref{eq:model_eqs}) by
means of the ADM and merely obtained the Taylor expansion of the solutions
about $t=0$:
\begin{equation}
x(t)=\sum_{j=0}^{\infty }x_{j}t^{j},\,y(t)=\sum_{j=0}^{\infty
}y_{j}t^{j},\,z(t)=\sum_{j=0}^{\infty }z_{j}t^{j}  \label{eq:xyz_series}
\end{equation}
where $x_{0}=x(0)$, $y_{0}=y(0)$, $z_{0}=z(0)$ are the initial conditions.
Obviously, if we substitute the series (\ref{eq:xyz_series}) into the
differential equations (\ref{eq:model_eqs}) we easily obtain the
coefficients $x_{j}$, $y_{j}$ and $z_{j}$ in terms of the initial conditions
and the model parameters without recourse to any elaborate method like the
ADM.

Later, Rafei et al applied the VIM\cite{RDG07} and the HPM\cite{RGD07} and
obtained the same Taylor series. In fact, the three papers show almost the
same partial sums of the series (\ref{eq:xyz_series}) and for that reason
display approximately the same figures for the discussion of the results\cite
{B06,RDG07,RGD07}.

The SIR model is the simplest description of the evolution of an epidemic of
a non--fatal disease. It does not consider birth or death of the individuals
and therefore the total population is constant
\begin{equation}
x(t)+y(t)+z(t)=x_{0}+y_{0}+z_{0}
\end{equation}
Besides, the model is so simple that it can be solved exactly and one
obtains
\begin{eqnarray}
y(x) &=&y_{0}+x_{0}-x+\frac{\gamma }{\beta }\ln \frac{x}{x_{0}}  \nonumber \\
z(x) &=&z_{0}-\frac{\gamma }{\beta }\ln \frac{x}{x_{0}}  \label{eq:y(x),z(x)}
\end{eqnarray}
However, if one needs the solutions in terms of $t$ one has to solve the
following integral numerically:
\begin{equation}
t=\frac{1}{\beta }\int_{x}^{x_{0}}\frac{dx^{\prime }}{x^{\prime }\left(
y_{0}+x_{0}-x^{\prime }+\frac{\gamma }{\beta }\ln \frac{x^{\prime }}{x_{0}}%
\right) }  \label{eq:t(u)}
\end{equation}

There are several important questions about the epidemic that one would like
to answer, for example, if we introduce a small number of infectives in a
population of susceptibles, will the number of infectives increase, causing
an epidemic, or will the disease fizzle out?. Assuming there is an epidemic,
how will it end?, will there still be susceptibles left when it is over?.
How long will the epidemic last?\cite{I08}. Biazar and Rafei et al\cite
{B06,RDG07,RGD07} merely showed some figures that are meaningless if they
did not answer any important question about the epidemic.

The exact solution enables one to answer the questions above. As an example
we analyze how the epidemic dies out. Notice that $dx/dt=dy/dt=dz/dt=0$ when
$y=0$, which according to the exact solution takes place when $x=x_{L}$ that
is a solution of
\begin{equation}
y_{0}+x_{0}-x_{L}+\frac{\gamma }{\beta }\ln \frac{x_{L}}{x_{0}}=0
\label{eq:x_L}
\end{equation}
We may say that the epidemic is over when the number of infectives is
similar to the one we had at the beginning: $y_{over}=y_{0}$; in this case we
have
\begin{equation}
x_{0}-x_{over}+\frac{\gamma }{\beta }\ln \frac{x_{over}}{x_{0}}=0
\label{eq:x_over}
\end{equation}

Biazar and Rafei et al\cite{B06,RDG07,RGD07} chose the following example
\begin{equation}
x_{0}=20,\,y_{0}=15,\,z_{0}=10,\,\beta =0.01,\,\gamma =0.02
\label{eq:mod_param_1}
\end{equation}
What we first appreciate here is that the number of infectives is rather
large: 1/3 of the total population. This situation does not define the
beginning of the epidemic but and advanced stage of it. However, we analyze
the results anyway.

\begin{figure}[H]
\begin{center}
\bigskip\bigskip\bigskip \includegraphics[width=9cm]{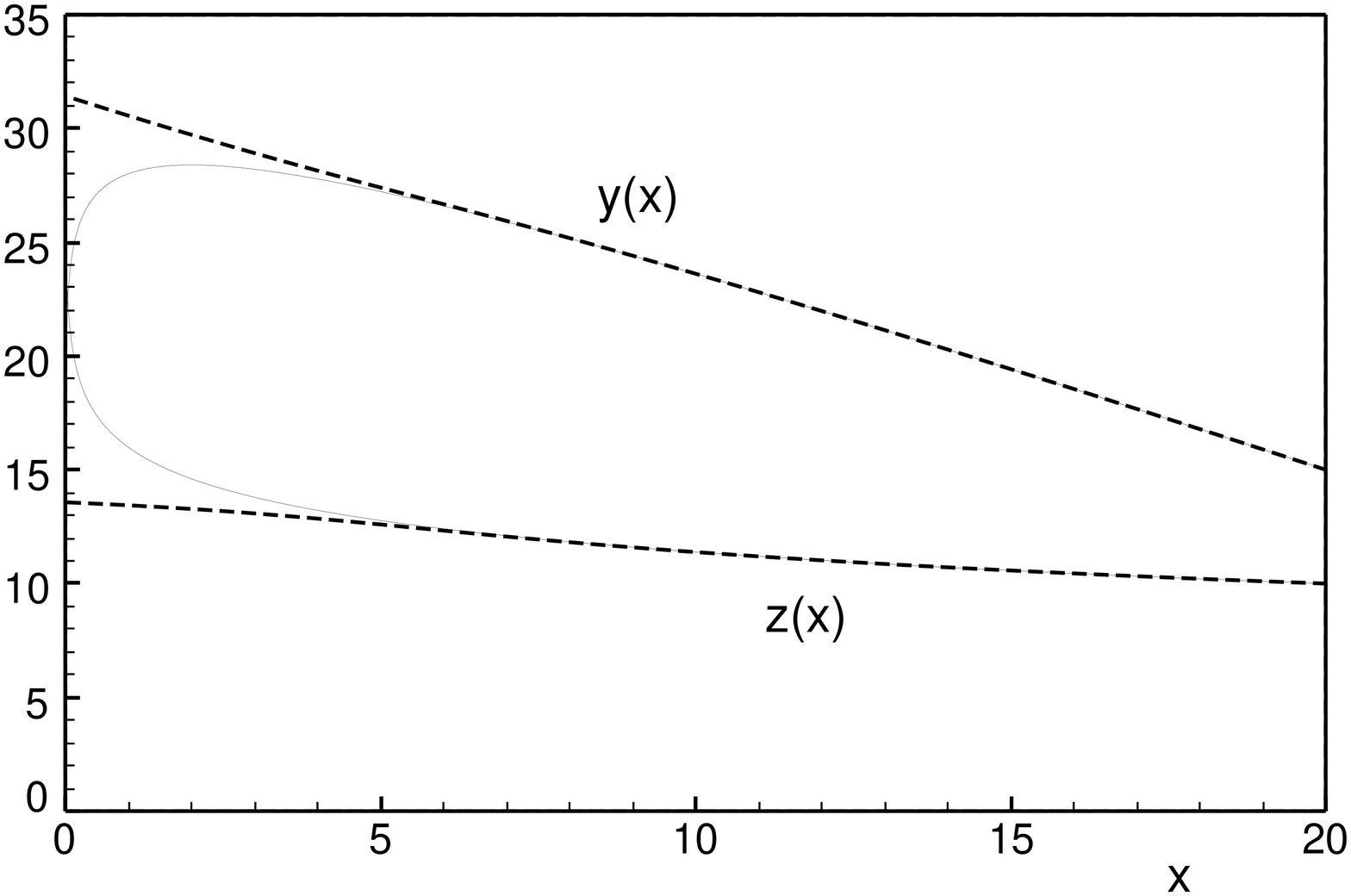}
\end{center}
\caption{Number of infectives $y$ and recovered $z$ in terms of the number
of susceptibles $x$ for the model parameters (\ref{eq:mod_param_1}). The
solid and dashed lines correspond to the exact and approximate results,
respectively.}
\label{fig:epi1}
\end{figure}

The fifth--degree polynomials proposed by Biazar and Rafei et al\cite
{B06,RGD07} do not give acceptable solutions for $y=0$ or $y=y_{0}$;
therefore, they cannot answer the relevant questions raised above. In order
to find the cause of the failure of the time--power series, we show $y(x)$
and $z(x)$ in Fig.~\ref{fig:epi1}. We appreciate that the time series do not
follow the exact curve in the region were we expect to find $%
x_{L}=5.02\times 10^{-7}$ and $x_{over}=0.000908$.

If $x_{0}>\gamma /\beta $ the number of infectives increases up to a maximum
at $x_{m}=\gamma /\beta $ and then decreases. The figures shown by Biazar
and Rafei et al\cite{B06,RDG07,RGD07} are restricted to a rather small time
interval and do not show this maximum. Besides, they suggest an ever
increasing number of infectives and do no reveal anything relevant about the
future of the dynamics of the epidemic process. For the particular model
parameters (\ref{eq:mod_param_1}) we obtain $x_{m}=2$ and $y_{m}=28.39$.

\begin{figure}[H]
\begin{center}
\bigskip\bigskip\bigskip \includegraphics[width=9cm]{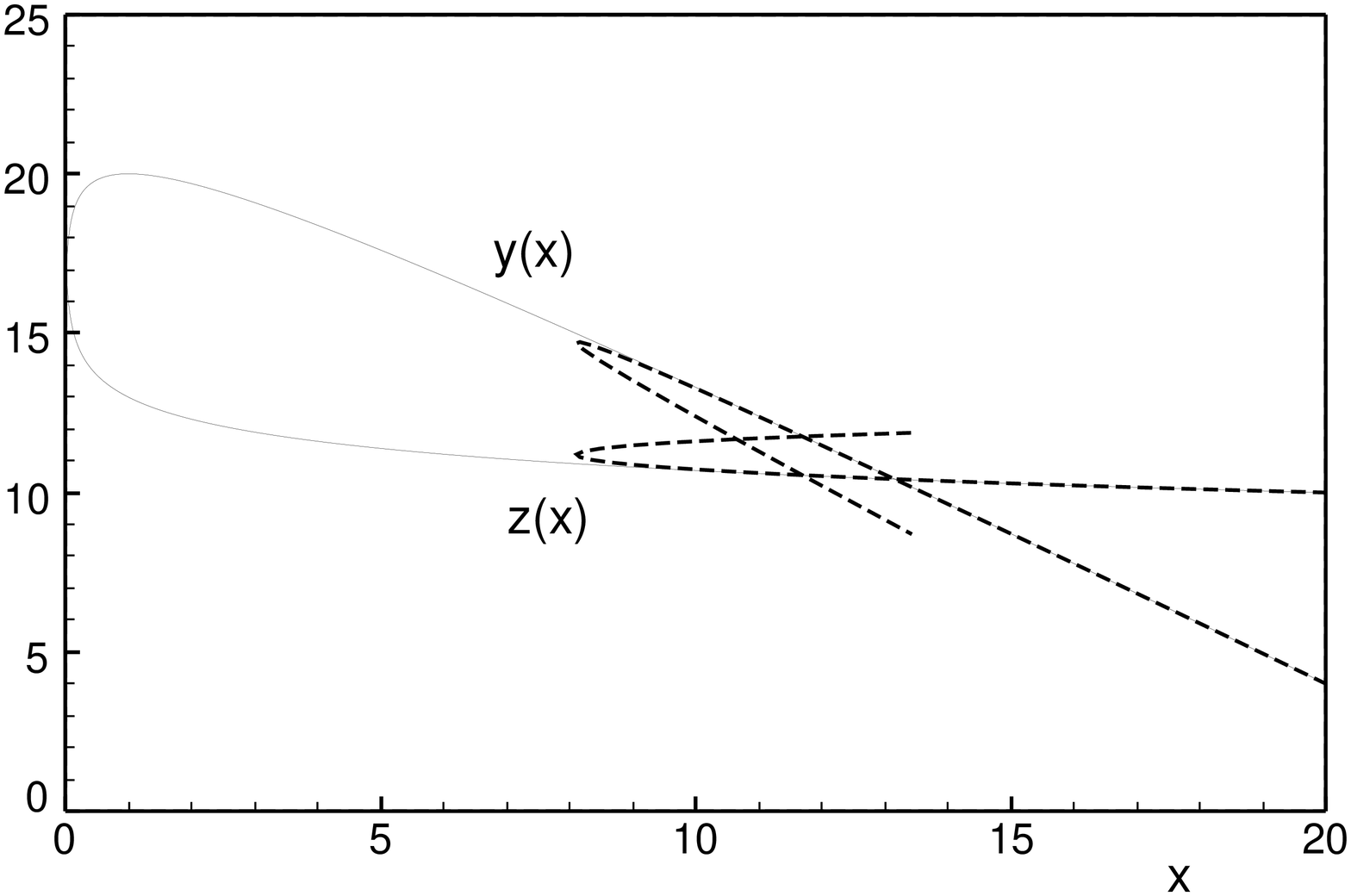}
\end{center}
\caption{Number of infectives $y$ and recovered $z$ in terms of the number
of susceptibles $x$ for the model parameters (\ref{eq:mod_param_2}). The
solid and dashed lines correspond to the exact and approximate results,
respectively.}
\label{fig:epi2}
\end{figure}

Although the power--series approach is unable to show the overall picture of
the epidemic dynamics it does not perform too poorly in the case of the
conveniently chosen model parameters (\ref{eq:mod_param_1}). One expects
that the range of utility of the time--power series decreases as the model
parameters $\beta $ and $\gamma $ increases. For example, if we choose
\begin{equation}
x_{0}=20,\,y_{0}=4,\,z_{0}=10,\,\beta =\gamma =1  \label{eq:mod_param_2}
\end{equation}
the situation is considerably worse than in the preceding case. Fig.~\ref
{fig:epi2} shows the exact curves $y(x)$ and $z(x)$ and those predicted by
the fifth--degree power--series approach. We appreciate that in this case
the time series fails to give any reasonable account of the epidemic
dynamics.

The results derived above clearly show that the ADM, VIM and HPM
implemented by Biazar\cite{B06} and Rafei et al\cite{RDG07,RGD07} are
unsuitable for the description of the epidemic dynamics. The reason is that
the time--power series is a local approximation valid in a relatively small
neighbourhood of the initial stages of the epidemic. Consequently, it cannot
provide the long--term behaviour of the infectious process that is needed
for understanding its future evolution. As we have already seen, the
analytical expressions provided by Biazar\cite{B06} and Rafei et al\cite
{RDG07,RGD07} do not allow us to answer the most relevant questions about
the epidemic dynamics, even in the case of an oversimplified model with
exact analytical solution. It is most probable that their performance may be
even poorer in the case of a more elaborated model.

In conclusion, we do not recommend a health--care system to rely on the ADM,
VIM or HPM time--power series to cope with an actual epidemic emergency.

\section{Conclusions}\label{sec:conclusions}
In our opinion most of the recent applications of VAPA are responsible for the
poorest scientific papers ever published. They are spreading like an epidemic
overcoming the refereeing mechanism of the journals. I believe that the main
reason for it is that the authors of such papers referee themselves and accept
the manuscripts that give them a considerable number of citations. Unfortunately,
it seems that some editors are accomplices of this situation favouring such
papers and banning criticisms of them. For example, some editors think that a
prey--predator model that predicts a negative number of rabbits is a valuable
scientific contribution to the journal\cite{F08d}. The reader may find some
more examples elsewhere\cite{F07, F08b,F08,F08c,F08d,F08e,F08f}.

\end{document}